\documentclass[aps,prb,twocolumn,floatfix]{revtex4}

\usepackage{amssymb}
\usepackage{amsbsy}
\usepackage{amsmath}
\usepackage{epsfig}
\usepackage{graphics}
\usepackage{graphicx}

\begin{document}


\title{Exotic Freezing of Response in Quantum Many-Body System} 

\author{Arnab Das}

\affiliation{
The Abdus Salam International Centre for Theoretical Physics (ICTP),
Trieste 34151, Italy and \\
Theoretical Division (T-4), LANL, MS-B213, Los Alamos, 
NM 87545, USA
}
\date{\today}

\begin{abstract}
We show that when a quantum many-body system is subjected to
coherent periodic driving, the response may exhibit exotic freezing behavior
in high driving frequency ($\omega$) regime. 
In a periodically driven classical thermodynamic system, 
freezing at high $\omega$ occurs
when $1/\omega$ is much smaller than the characteristic relaxation time
of the system, and hence the freezing always increases there as $\omega$ is
increased. Here, in the contrary, we see surprising non-monotonic freezing
behavior of the response with $\omega$, showing curious peak-valley structure.
Quite interestingly, the entire system tends to freeze almost absolutely 
(the freezing peaks) when driven with a certain combination of  
driving parameters values (amplitude and $\omega$) due to coherent suppression
of dynamics of the quantum many-body modes, which has no classical analog.
We demonstrate this new freezing phenomenon analytically (supported by large-scale
numerics) for a general class of integrable quantum spin systems.

\end{abstract}

\maketitle

The field of driven dynamics in quantum 
many-body system has attracted a 
lot of theoretical
attention in last few years
(see, e.g., \cite{Ad-prb} - \cite{QQ-book}). 
The major part of this research concentrated 
mainly around slow quenching dynamics across 
quantum critical points (lines/surfaces) 
resulting in quantum Kibble-Zurek Mechanism (KZM)
\cite{Damski} - \cite{Dziarmaga} 
of generating scaling laws for
defect densities - a direct translation 
of the consequences of robust classical KZM\cite{Zurek} in the quantum regime.
The distinctive role of quantum coherence in 
driven many-body dynamics thus somehow 
remained still largely unexplored  
(see, however, \cite{Amit,Ortiz,Miyashita,hprl}).
Experimentally, on the other hand, this coherent regime is becoming 
much more accessible
in recent years, thanks to the breakthrough in realizing highly 
isolated many-body quantum systems
with long coherence time within the setup of cold atom in optical lattice  
(see e.g., \cite{Damski-Aphys}). 

In this paper we report an early attempt to explore this regime, studying the 
Schr\"{o}dinger dynamics of a class of integrable quantum spin systems.  
The study reveals two generic regimes of the driving frequency $\omega$:
In the large $\omega$ regime (defined later), 
we observe surprising non-monotonic behavior of the response with respect to $\omega$, 
showing peak and valley structures, which dramatically 
contrasts the expected monotonic behavior ubiquitous in the corresponding 
classical scenarios of periodic driving (say, where a classical magnet is driven 
externally by a time-periodic magnetic field at finite $T$)
\cite{bkc-rmp,ma,tome,Zurek}. 
Here, for certain combinations of driving parameters,
($\omega$ and the driving amplitude $h_{0}$) 
the entire many-body dynamics freezes almost absolutely
giving rise to spectacular peaks. 
In the low $\omega$ regime, however, the peaks smooth out and a (roughly) monotonic behavior 
emerges as expected. We illustrate the crucial role of quantum coherence behind the phenomenon.
To clarify the analogy, we may note that in a driven classical thermodynamic system, 
``following the driving field" means following the trail of instantaneous thermal equilibrium states
corresponding to the instantaneous values of the time-varying driving field, 
where as for a quantum system at $T=0$, this means
following the instantaneous ground state of the time-varying Hamiltonian.   

In the classical case, faster driving always tends to leave the response more frozen in a  
monotonic manner in the high $\omega$ regime. The rationale is:
a faster driving would allow lesser time for the system to react
and hence the response would be left more frozen. 
The most universal and successful theory for non-equilibrium response behavior
(e.g., defect formation) in a driven classical system, namely the KZM, is based
on this rationale \cite{Zurek}. 
The quantum version of KZM \cite{Damski} - \cite{Dziarmaga} is also based on
the same classical notion of freezing - the response of a driven system gets
``frozen" when its instantaneous relaxation rate falls below the driving rate.
The response therefore remains frozen over a finite region around the critical point, where
this condition is met. 
The faster the driving rate is, the larger is this region of freezing
and thus more frozen is the response (e.g., for linear driving across a critical point, 
the said region of freezing increases as a power of
the driving rate, the power being given by the static critical exponents 
\cite{Zurek,Dziarmaga}).  
 
\begin{figure}[htb]
\begin{center}
%
\includegraphics[width=0.45\linewidth,height=0.4\linewidth]{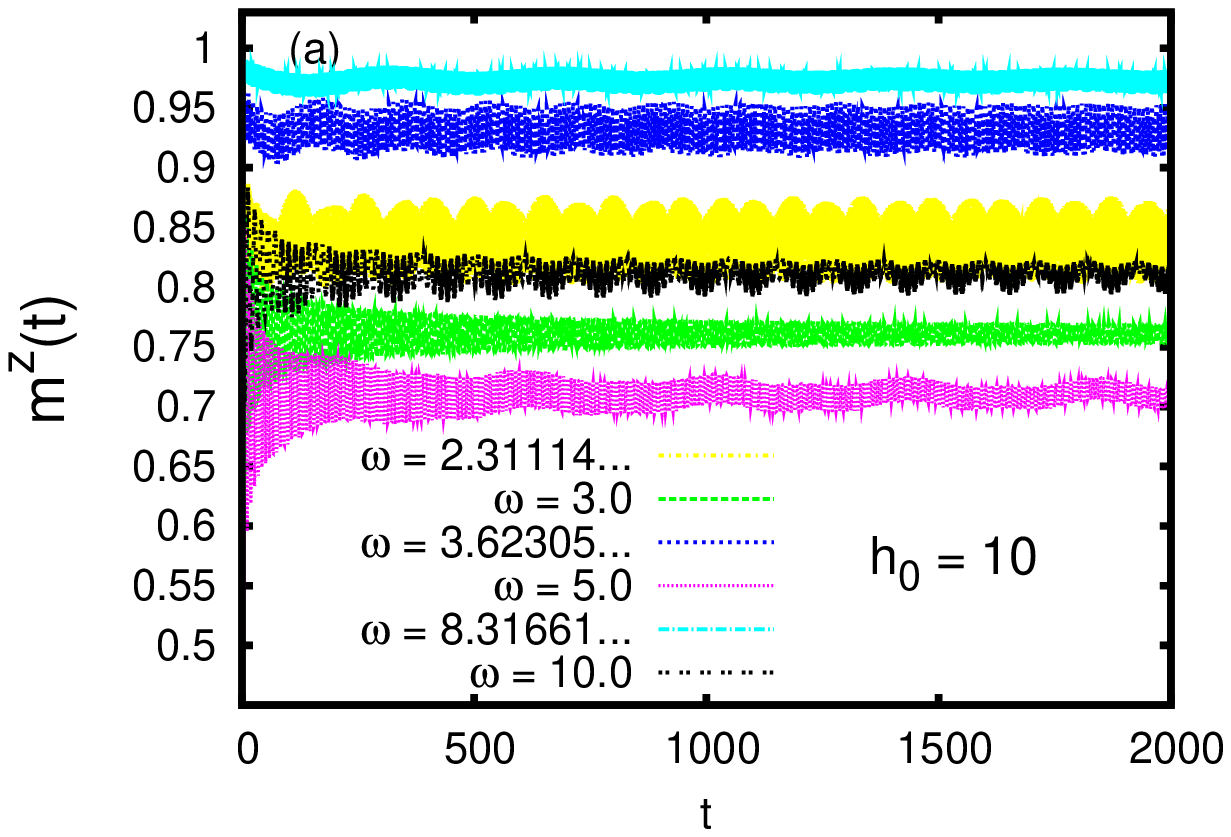}
\includegraphics[width=0.45\linewidth,height=0.4\linewidth,angle=0]{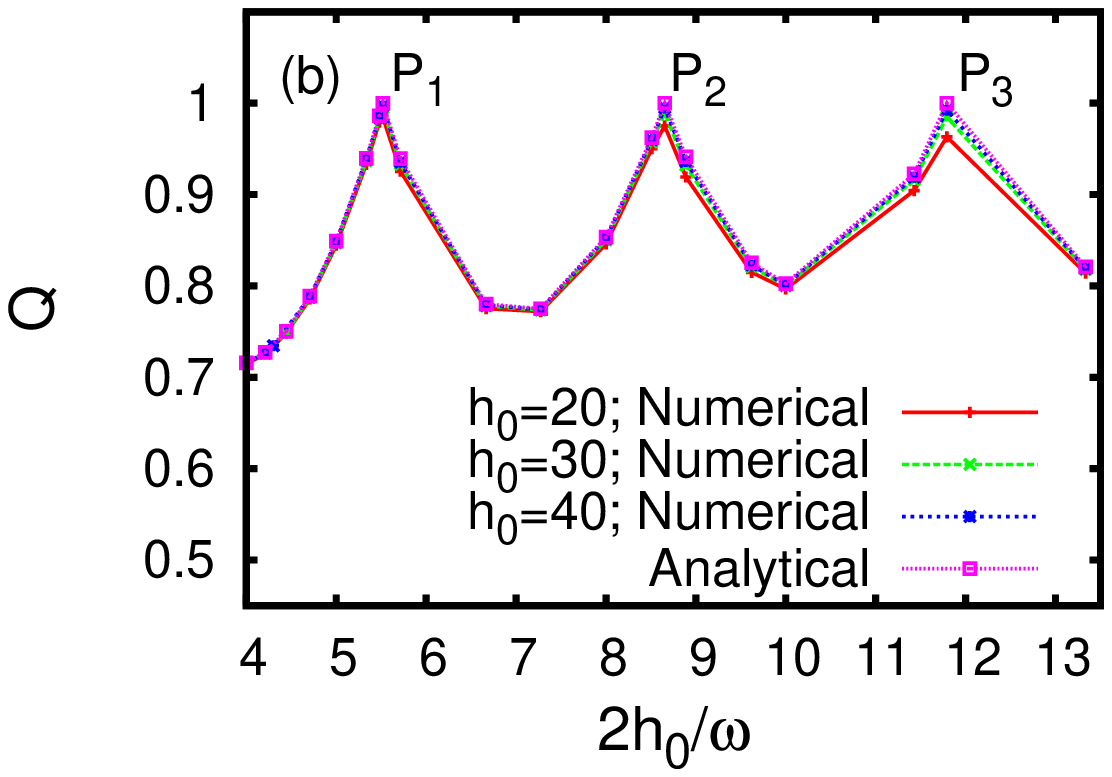}
\includegraphics[width=0.45\linewidth,height=0.4\linewidth]{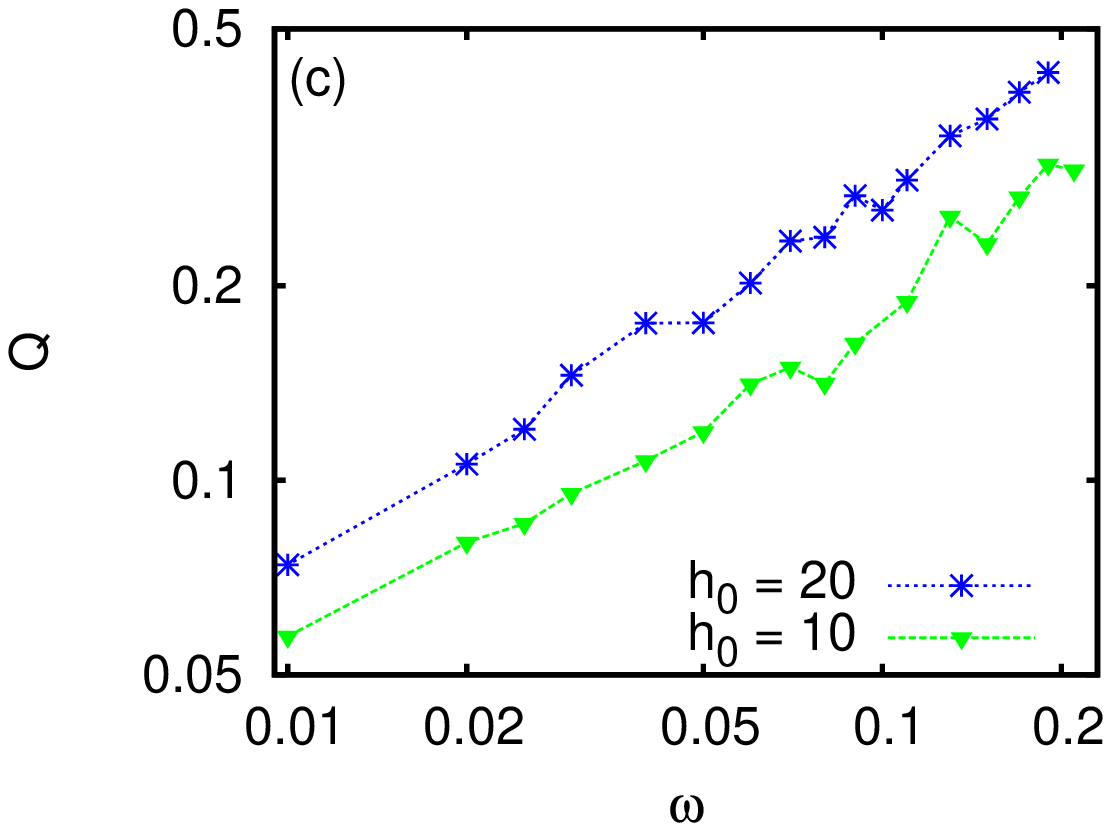}
\includegraphics[width=0.45\linewidth,height=0.4\linewidth]{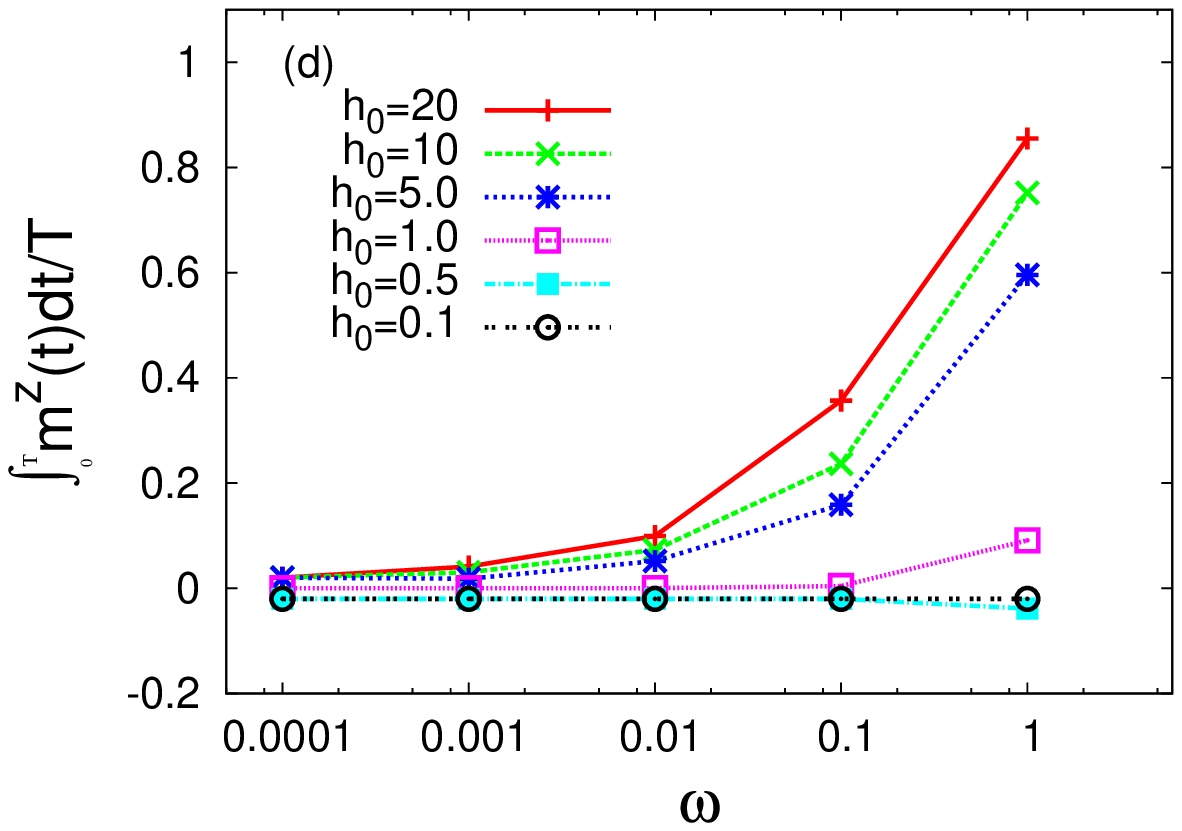}
%
\end{center}
\caption{\footnotesize{
(color online) {\bf (a)} $m^{z}$ Vs $t$ (numerical, $N=10^{4}$,$h_{0}=10$)
for various $\omega$s.
{\bf (b)} $Q$ Vs  $2h_{0}/\omega$ for $h_{0} =$ 20, 30 and 40, compared
with the analytical formula $Q = \frac{1}{1+|J_{0}(2h_{0}/\omega)|}$ 
(Eq. \ref{m-avg-analytical}).  
The peaks $P_{1}$, $P_{2}$
and $P_{3}$ representing maximal freezing, corresponds to three zeros of 
$J_{0}(\frac{2h_{0}}{\omega})$, occurring at 
$\frac{2h_{0}}{\omega} = 5.520..., 8.653...$ 11.971... respectively. 
{\bf (c)}  
$Q$ Vs $h_{0}$ for $\omega \ll 1$ 
(numerical). The peaks are smoothed out yielding a roughly monotonic behavior. 
{\bf (d)} The counter-classical trend of stronger freezing
for higher $h_{0}$
is demonstrated with single-sweep results;
we plot $Q_{s} = \frac{1}{T}\int_{0}^{T} m^{z}(t)dt$,
 (where $T = 2\pi/\omega$) for $N=100$. 
The Fig. also demonstrates monotonic
adiabatic to non-adiabatic transition 
at low $\omega$ regime for different $h_{0}$. 
}}
\label{Fig-2}
\end{figure}

Though the above classical rationale leads to 
the correct trend of the freezing (the scaling-law) 
even for the quantum systems 
in the small $\omega$ regime, here 
we demonstrate that it may surprisingly fail in some other cases when $\omega$ is high enough. 
In this regime, additional
freezing  may occur due to dynamics-dependent coherent cancellation of transition amplitude. 
We derive closed-form analytical expression for the entire
non-monotonic profile of the response, 
which accurately reproduces the 
(directly integrated) numerical results  
for large system-size ($N = 10^4$).    
The dependence of the response on the amplitude of the driving field
is also shown to exhibit trend completely
reverse of that observed with incoherent
classical fluctuations. 
We demonstrate this quantum freezing phenomena 
for a general class of integrable $d-$dimensional quantum system with Hamiltonians
of the form (in momentum space)
\begin{equation}
{\mathcal{H}}(t) =\sum_{\vec{k}} \psi_{\vec{k}}^{\dagger} 
\left(
\begin{array}{cc}
h_{z}(t)+f_{\vec{k}} & \Delta_{\vec{k}}\\
\Delta_{\vec{k}}^{\ast} & -h_{z}(t)-f_{\vec{k}}
\end{array}
\right)
\psi_{\vec{k}}
\label{H-d}
\end{equation}
\noindent 
where $\psi_{\vec{k}} = (c_{1\vec{k}}, c_{2\vec{k}})$ are standard fermionic operators
in $\vec{k}$-space, 
$h_{z}(t) = h_{0}\cos{(\omega t)}$ 
is the driving field (any Hamiltonian parameter) and 
$f_{\vec{k}}$ (real) and $\Delta_{\vec{k}}$ are system-specific functions.  
The above-mentioned class includes many well known quantum spin models in one, two and three
dimensions, such as the transverse field Ising model (TFIM), quantum X-Y model,
and extended Kitaev models (see \cite{Kris}$^{(b)}$,\cite{kitaev} and references therein). 
We focus on TFIM for concrete illustration of the phenomenon because of its
intuitive appeal 
but keep the calculation general so that the main result is
easily visible for all the above-mentioned models.
For TFIM with Hamiltonian
\begin{equation}
H(t) = -\frac{1}{2}\left[J\sum_{i=1}^{N} \sigma_{i}^{x}\sigma_{i+1}^{x}
+ h_{z}(t)\sum_{i=1}^{N} \sigma_{i}^{z}\right],
\label{H-Ising}
\end{equation}
\noindent
($\quad h_{z}(t)= h_{0}\cos{(\omega t)}$ 
is the driving field and
$\sigma^{x/z}_{i}$
are $x/z$ component of the Pauli spin),
the form in (\ref{H-d}) with  
$f_{\vec{k}} = J\cos{k}$ and $\Delta_{\vec{k}} = J\sin{k}$
is obtained via 
Jordan-Wigner transformation followed by
Fourier transform. 
If one starts with the ground state of the Hamiltonian at $t = 0$,
the time-dependent wave function $|\psi(t)\rangle$ for the
system may be expressed as a direct-product
of two-dimensional time-dependent wave functions:
\mbox{
$|\psi(t)\rangle = \bigotimes_{k>0}|\psi_{k}(t)\rangle$}
with
$
|\psi_{k}(t)\rangle = u_{\vec{k}}(t)|0_{+k}0_{-k}\rangle
+ v_{\vec{k}}(t)|+k,-k\rangle,
$
where $|0_{+k}0_{-k}\rangle$ and $|+k,-k\rangle$
represent respectively the unoccupied and
the doubly occupied states of the
$\pm k$-fermions.
We start at $t = 0$
from the ground state of the Hamiltonian $H(t=0)$ in
Eq. (\ref{H-Ising})
with $h_{0} \gg 1$
(highly polarized in $+z$-direction), and as
the field oscillates we
measure the corresponding response function,
the transverse magnetization:
\begin{equation}
m^{z}(t) = \langle\psi(t)|\sigma_{i}^{z}|\psi(t)\rangle
=  \frac{4}{N}\sum_{k>0} |v_{\vec{k}}(t)|^{2} - 1.
\label{Transmag}
\end{equation}
\noindent
The resulting time evolution of $m^{z}$, obtained by
numerical integration of the
time dependent Schr\"{o}dinger equation
(TDSE) for many sweeps, is shown in
Fig. \ref{Fig-2}(a), for different $\omega$'s.
In each case, $m^{z}$ is found to
remain confined within a narrow
range in the positive sector (starting with $m^{z} \approx 1$)
for all time, though the field $h_{z}$ oscillates symmetrically about zero.
The dynamical symmetry breaking (due to freezing near the initial state) 
is quantified
by the  so called dynamical order parameter $Q$, which is the long-time
average of the response function \cite{bkc-rmp}. For TFIM 
\begin{equation}
Q = \langle m^{z} \rangle =
\lim_{T_{f}\rightarrow \infty}\frac{1}{T_{f}}\int_{0}^{T_{f}}
m^{z}(t)dt.
\label{<mz>}
\end{equation}
\noindent
Dynamical symmetry breaking ($Q \ne 0$) is shown 
for different values of $\omega$
and $h_{0}$ in Fig. \ref{Fig-2}(a,b).

The intuitive reason for non-zero $Q$ might simply be the lack of adiabaticity
in the dynamics. 
Non-adiabaticity occurs when 
the characteristic response time of a quantum system, given by the
inverse of the relevant energy gap, is large compared to the driving period.
Thus freezing due to the competition 
of these two timescales 
(the basis of quantum KZM estimate 
\cite{Damski}-\cite{Dziarmaga}) 
is similar in spirit 
(though very different in mechanism)
with the classical dynamical hysteresis
(freezing of the magnetization dynamics in a classical magnet when driven 
too fast by a periodic magnetic field). The similarity is only in the sense, 
that both freezing represent the failure of the system's reflex
to adjust to the rapidly changing field, and 
hence in both cases
stronger freezing is expected as the driving is made faster. 
In the classical case, the freezing is in fact always 
monotonic with respect to $\omega$.  
For example, 
in case of a periodically driven classical Ising model\cite{bkc-rmp},
the frozen (asymmetric; $Q\ne 0$) phase shrinks monotonically 
in the symmetric-asymmetric phase diagram as $\omega$ is reduced and 
finally vanishes as $\omega \rightarrow 0$.
This monotonicity is also observed even here, 
when $\omega$ is small enough, as shown in Fig. \ref{Fig-2}(d). 
Surprisingly, 
contrary to this picture, we find
$Q$
to be a non-monotonic function
of $\omega$, exhibiting peaks appearing 
at high $\omega$'s
as shown in Fig. \ref{Fig-2}(b).
The peaks represent maximal freezing of the system
with $Q$ very close to unity,
indicating additional freezing for
certain combinations of $\omega$ and $h_{0}$. 
In the low frequency regime, however, the peaks 
are found to be smoothed out (roughly), and 
$Q$ is observed to 
decrease more or less monotonically with $\omega$
as shown in Fig. \ref{Fig-2}(c).
In analogy with the classical case, we call this non-monotonic quantum freezing phenomenon
(and its related dynamical symmetry breaking) dynamical quantum hysteresis (DQH). 
  
We explain the above scenario solving
the dynamics of the $k-$modes as follows.
Employing the $2\times 2$ unitary transformation
${\hat U}_{k} = \exp\left[-\frac{i}{2}(2tf_{\vec{k}}
+ \frac{2h_{0}}{\omega}\sin{\omega t})\sigma^{z} \right]$
on $|\psi_{k}(t)\rangle$ and $H_{k}(t)$
and performing a subsequent expansion in terms of Bessel's functions:
$\exp{\left[iz\sin{\theta}\right]}
= \sum_{n=-\infty}^{+\infty} J_{n}(z)e^{in\theta}$ 
($J_{n}(z)$ being the Bessel's function of
the first kind with integer order $n$) 
\cite{ashhab},
we get
the transformed wave function
$
|\psi_{k}^{\prime}(t)\rangle
=u_{\vec{k}}^{\prime}(t)|0_{+k},0_{-k}\rangle + v_{\vec{k}}^{\prime}(t)|+k,-k\rangle
\label{psi-k-prime}
$
that follows the TDSE
with the transformed Hamiltonian:
\begin{equation}
H^{\prime}_{k}(t) = -\Delta_{\vec{k}}
\left(
\begin{array}{cc}
0 & i\sum_{n=-\infty}^{\infty}R_{n}\\
-i\sum_{n=-\infty}^{\infty}R_{n}^{\ast} & 0
\end{array}
\right)
\label{H-intrctn}
\end{equation}
\noindent
where 
$
R_{n} = J_{n}\left(\frac{2h_{0}}{\omega}\right)
e^{-i(n\omega + 2f_{\vec{k}})t}.
$
We note that
$|u_{\vec{k}}^{\prime}(t)|^{2} = |u_{\vec{k}}(t)|^{2}$ and
$|v_{\vec{k}}^{\prime}(t)|^{2} = |v_{\vec{k}}(t)|^{2}$.
Now we resort to the resonance approximation (RA),
under which the equation are soluble
(see, e.g.,\cite{silverman}).
The RA amounts to
ignoring all the (faster oscillating) terms
in the off-diagonal sum in $H^{\prime}_{k}(t)$
except for the resonant term $n = n_{r}$,
for which the effective frequency
$\Omega_{k} = |n\omega + 2f_{\vec{k}}|$
is the smallest.
In the high frequency limit
($\omega \gg 2|f_{\vec{k}}|$), we have $n_{r} = 0$.
Physically, this means, $\omega$ is far off-resonant with
the relevant characteristic frequencies of the system given by $2J\cos{k}$ ($J=1$ here).  
For the general d-dimensional Hamiltonian (\ref{H-d}), the criteria reads $\omega \gg |f_{\vec{k}}|$.
The general solution of the TDSE with Hamiltonian
(\ref{H-intrctn}) under RA gives
$
v_{\vec{k}}(t) = 
-2ie^{if_{\vec{k}}t}
\left[\frac{J_{0}(2h_{0}/\omega)\Delta_{\vec{k}}}
{2\phi_{\vec{k}}}\sin{(\phi_{\vec{k}}t)}\right]u_{\vec{k}}(0) \\ \nonumber
+
e^{if_{\vec{k}}t}\left[\cos{(\phi_{\vec{k}}t)} 
- i\frac{f_{\vec{k}}}{\phi_{\vec{k}}}\sin{(\phi_{\vec{k}}t)}\right]v_{\vec{k}}(0),
$
where $\phi_{\vec{k}} = \sqrt{J_{0}^{2}(2h_{0}/\omega)\Delta^{2}_{\vec{k}} + f^{2}_{\vec{k}}}$.
With the initial condition mentioned above
($|v_{\vec{k}}(0)|^2 \approx 1$),
one gets 
\begin{equation}
|v_{\vec{k}}(t)|^{2} = 1 -
\frac{J_{0}^{2}(2h_{0}/\omega)\Delta_{\vec{k}}^{2}}{J_{0}^{2}(2h_{0}/\omega)\Delta_{\vec{k}}^{2}
+ f^{2}_{\vec{k}}}\sin^{2}{(\phi_{\vec{k}}t)}.
\label{RA-Gen}
\end{equation}
\noindent
From Eq. (\ref{RA-Gen}) we see that for
the near-critical modes ($k \sim 0, \pi,\quad 
\Delta_{\vec{k}} \sim 0$),
$|v_{\vec{k}}(t)|^{2}$ oscillates with a vanishing
amplitude proportional to $\Delta_{\vec{k}}^2$,
and thus contribute maximally to the freezing.
On the other hand, the off-critical modes
($k \sim \pi/2$) undergo full oscillation without
any appreciable freezing.
The
intermediate modes,
($1 > |\Delta_{\vec{k}}| \gg 0$) 
oscillating with an amplitude that depend both on
$k$ and the ratio $h_{0}/\omega$ (Eq. \ref{RA-Gen}) contribute
non-trivially to the freezing. Non-monotonic freezing is
encoded here in the non-monotonicity of $J_{0}(x)$.
Any local observable is obtained by
summing up the contributions from all these non-local many-body
modes. 
To get an explicit formula for $Q$ for TFIM, 
we set $\Delta_{\vec{k}} = \sin{k}$,
$f_{\vec{k}} = \cos{k}$ and
take the continuum limit of Eq. (\ref{Transmag}).
Integrating over $k$, taking the limit
 $T_{f} \rightarrow \infty$
of Eq. (\ref{<mz>}) gives a
simple formula:
\begin{eqnarray}
Q &=& \frac{1}{1 + |J_{0}(2h_{0}/\omega)|.}
\label{m-avg-analytical}
\end{eqnarray}
\noindent
The expression matches remarkably well with the peaked-structured
profile of $Q$
obtained by numerical integration,
as shown in Fig. \ref{Fig-2}(b).
The peaks occur for certain combinations of $\omega$
and $h_{0}$, for which
$J_{0}\left(\frac{2h_0}{\omega}\right) = 0$. Under this
condition, all the modes freeze, resulting in
an absolute localization (within the RA approximation made) of the system at its initial
state for all time (known as coherent destruction of tunneling
in the context of driven two-level system \cite{hanggi}). 
The exact form of $Q$ depends on the model dimension and other
system-specific details, but the key feature - non-monotonicity
is already reflected in the general equation Eq.(\ref{RA-Gen}).

To clarify the fundamental difference between the 
nature of this additional freezing
and the freezing due to KZM,
we note that 
after the first sweep
the non-adiabatic excitation
probability 
$p_{k} = |v_{k}(\omega t=\pi)|^{2}$ 
for each mode
(related both to the total defect-density\cite{Amit}
and $m^{z}$, Eq. \ref{Transmag}), 
is actually a non-monotonic function 
of $\omega$ (Eq.\ref{RA-Gen}). 
In contrast, KZM
would predict in such cases, a
monotonic increase of the size of the 
impluse-region (2$\hat{\epsilon}$)
with $\omega$, 
resulting in
a monotonically increasing $p_{k}$  
(see, e.g., \cite{Dziarmaga}). This monotonicity
is a general charactersistic of KZM as long as
$|\dot{\epsilon}|$ is either constant
or increases monotonically as the critical point is
approached\cite{Zurek}.


In the regime  $\omega \ll |f_{\vec{k}}|$, however, the
off-diagonal sum in $H^{\prime}_{k}(t)$ cannot be approximated by 
a single-frequency term. 
Presence of many close multiples of $\omega$
satisfying resonance condition
$\Omega_{k} = |n\omega + 2f_{\vec{k}}|  \sim 0$, 
smooths out the peaks and a gross monotonic behavior emerges, 
as shown in Fig. \ref{Fig-2}(c). The dynamics remain non-adiabatic
due to  quantum critical
points at $h_{z} = \pm 1$ for any nonzero 
$\omega$ in the thermodynamic limit. 
A more detailed study of the
low-frequency behavior will be reported elsewhere 
\cite{AD-Project-2}. 

Phase coherence plays a crucial
role in determining $Q$,
as can be 
seen from the following example. 
In the limit $h_{0} \gg |\sin{(k)}|$, the 
evolution corresponding to a full driving cycle
might be decomposed into adiabatic and impulse regimes \cite{Damski,ashhab},
such that apart from some neighborhood $\pm \Delta h$ of the critical points
at $h_{c} = \pm 1$, the dynamics is adiabatic, while within these neighborhoods
the dynamics is impulsive, and can be approximated by 
Landau-Zener transitions 
upon linearizing the sinusoidal field for low enough $\omega$.  
Now, if the phase-coherence between the fermionic state $|0_{+k},0_{-k}\rangle$ 
and $ v_{\vec{k}}^{\prime}(t)|+k,-k\rangle$ is neglected (see for example, \cite{Amit-deco}), 
say, due to some decoherence
mechanism, 
then the fermionic excitations  
after $n$ complete
cycles would be given by 
$ |v_{\vec{k}}(nT)|^2 = \frac{1}{2}[1 + (2\theta_{k} - 1)^n]$, where 
$\theta_{k} = \exp{\left[-\frac{\pi\sin^2{(k)}}{\omega\sqrt{h_{0}^2 -\cos^2{k}}}\right]}$. This
implies, $m^{z}$ approaches $0$ rapidly, giving $Q = 0 \forall \omega$, contrasting
the coherent case results (Fig. 1b).

The behavior of $Q$ with $h_{0}$ also
 contrasts the classical picture in a drastic way.
In the classical case of a periodically driven magnet in
presence of thermal fluctuations,
dynamical localization always occurs
below a certain value of the amplitude $h_{0}$
(for a given $\omega$ and temperature),
above which the symmetric phase appears
 \cite{tome,bkc-rmp}. High enough driving fields
in a classical Ising magnet (even in quantum magnets
with some coherence)
kills any hysteresis/freezing,
forcing the system more strongly to follow the field
as demonstrated experimentally by Aeppli's group \cite{aeppli}.
But in DQH, just the reverse trend
is observed, as shown in Fig. \ref{Fig-2}(d) (low $\omega$ regime). 
In the high $\omega$ limit also, one has
$J_{0}(x) \approx \sqrt{\frac{2}{\pi x}}\cos{(x - \pi/4)}$
for $x \gg \frac{1}{4}$ and
the expression (\ref{m-avg-analytical})
reduces to
$Q \approx
1 - \frac{\sqrt{\omega}\cos{\left(\frac{2h_{0}}{\omega}-\frac{\pi}{4}\right)}}{\sqrt{\pi h_{0}}
+ \sqrt{\omega}\cos{\left(\frac{2h_{0}}{\omega}-\frac{\pi}{4}\right)}},
$
with $\lim_{h_{0}\rightarrow\infty}
Q = 1$, giving absolute freezing
of the dynamics in this limit.
A general qualitative explanation
of this reverse trend goes as follows.
In dynamics driven by classical
fluctuations, a stronger field
would induce stronger asymmetry between the rate of the
aligning (spins orienting parallel to the field)
and the anti-aligning dynamics, favoring the former
one energetically over the latter.
Hence for a higher $h_{0}$,
the spins would re-align faster along the field 
following a field reversal and thus the hysteresis/freezing would
be reduced. But in the case of coherent quantum
fluctuations, a stronger field would instead, 
more strongly suppress all the dynamics that
would change the response, 
even if it helps lowering the
field-induced potential enery,  
since the response (by definition) 
commutes with the
field part of the Hamiltonian.

Experimental observation of the DQH phenomenon
may be realizable in several ways. First, realization 
of this phenomenon would be possible in tunable transverse
Ising model using trapped ions \cite{friedenauer}.
A similar realization would be possible
within lattice-spin models with polar
molecules on optical lattices \cite{zoller}. In these systems,
the exchange interaction $J$
have experimental upper-limits
($\sim$ $22.1$ kHz and between $10 - 100$ kHz
for the respective cases mentioned above) 
but can be made arbitrarily small.
Hence the range of high $\omega$ referred here
(in the units of $J$) may be brought down to
a comfortable range, say to the order of few kHz in
both the realizations. 
The phenomenon of DQH in
Kitaev models \cite{kitaev}
can be achieved via experimental set-ups as proposed in \cite{duan}. 

Its implication in the context of 
quantum annealing \cite{ad-rmp}, 
might be quite interesting, as it may contrast
the intuitive scenario of monotonic improvement with 
slower annealing in certain cases. 

\vspace{0.16cm}
\noindent
{\bf Acknowledgments:} The author thanks B. K. Chakrabarti, 
G. Santoro, E. Tosatti, K. Sengupta, W. H. Zurek, B. Damski, 
N. Surendran and A. Silva for useful discussions.



\end{document}